\documentclass[preprint,amsmath,amssymb,aps,prl,showpacs,footinbib]{revtex4-1}
\usepackage{hyperref,verbatim,nicefrac,longtable}
\usepackage{braket,lineno,color,soul,graphicx}

\newcommand{\RE}{\operatorname{Re} }
\newcommand{\IM}{\operatorname{Im} }

\begin{document}
%TC:ignore
\title{Is the Vlasov equation valid for Yukawa plasmas?}
\author{Kenneth I. Golden$^1$, Gabor J. Kalman$^2$, Luciano G. Silvestri$^2$}
\affiliation{$^1$ Department of Mathematics and Statistics, Department of Physics, University of Vermont, Burlington, VT 05401, USA}
\affiliation{$^2$ Department of Physics, Boston College, Chestnut Hill, MA 02467, USA}

\begin{abstract}
We analyze the Vlasov dispersion relation for Yukawa plasmas in three
dimensions for the purpose of identifying coupling parameter domains where
the Vlasov approach is justified and the existence of a well-developed RPA type collective excitation is allowed. We establish a rigorous lower bound for the coupling parameter, below which there can be no real solution to the Vlasov dispersion relation. In the coupling domain, where weakly damped solutions do exist, we have focused on the long-wavelength acoustic regime
where we establish more restrictive estimates for the lower bound of the coupling
parameter. We also derive a general formula for the corresponding acoustic phase
velocity, valid over a wide range of coupling parameter/screening parameter ratios above
the lower bound. We conclude that the Vlasov approach is tenable only above a critical coupling value.
Comparison with Molecular Dynamics simulation results further highlights the limitations of the Vlasov approximation for weakly coupled Yukawa plasmas.
\end{abstract}
\date{\today}

\pacs{}
%TC:endignore
\maketitle
The collective behavior of a conventional weakly coupled Coulomb  plasma is well described by the Vlasov equation. The Vlasov kinetic equation can be used with relative ease for the calculation of the plasma dielectric function $\varepsilon(\mathbf k,\omega)$, which then can be invoked to determine the collective mode structure \cite{Ichimaru2004}. For the simple model of the one component plasma (OCP) the well-known conclusion is the existence of the plasma oscillation as the longitudinal collective mode, governed by the plasma frequency $\omega_p$ (defined below), whose dispersion $\omega(k) $ is given by the Bohm-Gross dispersion relation. This approach, however, has its limitations: once the value of the coupling parameter $\Gamma$ (defined below) exceeds 1, the plasma becomes strongly coupled and the dielectric response function shows a substantial deviation from its Vlasov structure, with a concomitant dramatic change of the plasmon dispersion.

Interest in multicomponent systems, where the basic Coulomb interaction between the dominant particles is screened by other charged particles in the system has led to the Yukawa plasma model, consisting of one single species of particles, whose interaction is through a short-range Yukawa potential with a screening parameter $\kappa$ (see eq.~\eqref{yukawa_r_potential}), rather than through the long-range Coulomb potential. The most prominent representatives of such systems are dusty plasmas.  Basic symmetry principles \cite{Landau2013}, supported by experimental observations \cite{DAngelo1990,Pieper1996,Thompson1997,Merlino1998,Nunomura2002}, demand that the longitudinal collective excitation of the Yukawa OCP  (YOCP) be an acoustic mode, \textit{i.e.} $\omega(k\rightarrow0) = sk$, where $s$ is the sound velocity. The YOCP may also be weakly or strongly coupled, determined by the value of its coupling parameter $\Gamma_\kappa$ (which, in general, is different from $\Gamma$). The reliable calculation of the sound velocity requires the determination of $\varepsilon(\mathbf k,\omega)$ and its employment to obtain the collective mode dispersion, in a fashion similar to what has been established  for the Coulomb OCP. The question then arises whether for a weakly coupled system the formalism that is based on the Vlasov equation may be used for the YOCP as well. In the existent literature on the subject (see \textit{e.g.} \cite{Rosenberg1993} ), the tacit assumption is made that the answer is in the affirmative.  The validity of the Vlasov equation, however, hinges upon two conditions: \textit{(i)} the long range character of the interaction, and \textit{(ii)} weak coupling \cite{Rosenbluth1959}. While the second condition may be satisfied, the first one is obviously not. Thus, we contend, the issue has to be carefully re-examined. This is the purpose of the present paper.  We will show that apart from the weak coupling requirement, $\Gamma_\kappa$ or $\Gamma$ are subject to a limitation, which restricts the validity of the Vlasov approach to a certain domain. We will also demonstrate that recent Molecular Dynamics (MD) simulations mostly corroborate this conclusion.

The Yukawa interaction potential has the form
\begin{equation}
\phi(r) = Q^2 \frac{ e^{-\kappa r} }{r}
\label{yukawa_r_potential}
\end{equation} 
with Fourier transforms
\begin{equation}
\phi(k) = \frac{4\pi Q^2}{\kappa^2 + k^2}, 
\label{3Dphik}
\end{equation}
$Q$ representing the charge of the particle and $\kappa$ the screening parameter. The coupling constant is defined as 
\begin{equation}
\Gamma = \beta\frac{Q^2}{a},
\label{coupling_const}
\end{equation}
where $\beta = 1/k_BT$ and $(4\pi/3) a^3n = 1$ defines the Wigner-Seitz radius $a$ in three dimensions. For $\Gamma \ll 1$ the system is considered weakly coupled while for $\Gamma > 10$ particles correlations become dominant and the system is in a strongly coupled phase and for $\Gamma > \Gamma_m$ the system forms a Wigner crystal. The exact value of $\Gamma_m$ and the crystal structure depend on $\kappa$ \cite{Hamaguchi1997}. The collective mode spectrum of the YOCP is governed by a single longitudinal acoustic mode that in the $\kappa \rightarrow 0$ limit morphs, mirroring the Anderson mechanism, into a longitudinal plasmon \cite{Anderson1963}. In contrast to the OCP, where the longitudinal plasmon, protected by the Kohn sum rule, is independent of correlations \cite{Brout1959}, the acoustic mode in the YOCP strongly depends on the coupling parameter. Focusing on the weak coupling domain, the aim of this paper is to see to what extend the Vlasov approach can be justified here and to find a bound in terms of $\Gamma,\kappa$ for the existence of a RPA type collective mode in the weak coupling regime. 

The collective mode is the solution of the dispersion relation \cite{Ichimaru1986}
\begin{equation}
\varepsilon(\mathbf k,\omega) = 1 - \phi(k)\chi(\mathbf k,\omega) = 0,
\label{epsilon}
\end{equation}
where $\chi(\mathbf k,\omega)$ is the screened (total) density response function,
\begin{equation}
\chi(\mathbf k,\omega) = - \frac{1}{m} \int d\mathbf v \frac{1}{\omega - \mathbf {k\cdot v}} \frac{\partial f(v)}{\partial \mathbf v},
\label{chi0}
\end{equation}
and $f(v)$ is the Maxwellian distribution function normalized to the average particle density $n$. On the premise that the Landau damping is weak, seeking the zeros of \eqref{epsilon} amounts to solving 
\begin{equation}
\frac{1}{\phi(k) }  = \RE \chi \left (\mathbf k,\omega(k) \right )
\label{phiRechi}
\end{equation}
for the (real) oscillation frequency $\omega(k)$, accompanied by the calculation of the companion damping rate
\begin{equation}
\gamma(k) = - \frac{ \IM \varepsilon \left( \mathbf k,\omega(k) \right )}{  \nicefrac{\partial }{\partial \omega} \RE \varepsilon \left(\mathbf k,\omega \right ) \left. \right |_{\omega = \omega(k)} } = 
 - \frac{ \IM \chi \left( \mathbf k,\omega(k) \right )}{  \nicefrac{\partial }{\partial \omega} \RE \chi \left(\mathbf k,\omega \right ) \left. \right |_{\omega = \omega(k)} } < 0.
\label{damping}
\end{equation}
In deriving eqs.~\eqref{phiRechi} and \eqref{damping} we have invoked the well-known weak-damping hypothesis 
\begin{equation}
\left | \IM \chi \left(\mathbf k,\omega(k) \right ) \right | \ll \left | \RE \chi \left( \mathbf k,\omega(k) \right ) \right |, \quad \left | \gamma(k) \right | \ll \omega(k).
\label{smalldamphypothesis}
\end{equation}

\begin{figure}[ht]
\centering
\includegraphics[width = .9\textwidth]{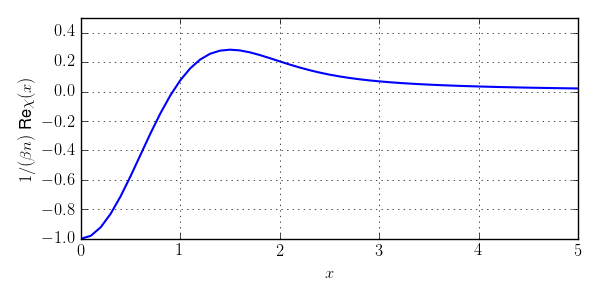}
\caption{Plot of $(1/\beta n) \RE \chi(x)$ as a function of $x = (\omega/k) \sqrt{\beta m/2}$.}
\label{fig:rechi}
\end{figure}

The following dispersion relation then results from eq.~\eqref{3Dphik}:
\begin{equation}
\frac{\bar\kappa ^2 + \bar k^2}{3\Gamma} = \frac{1}{\beta n} \RE \chi(x),
\label{3d-disp}
\end{equation}
\begin{equation}
\frac{1}{\beta n} \RE \chi(x) = -1 + \frac{x}{\sqrt{\pi} } P \int_{-\infty}^{\infty} dt \frac{e^{-t^2} } {x - t},
\label{rechi}
\end{equation}
\begin{equation}
\frac{1}{\beta n} \IM \chi(x) = -\sqrt{\pi} e^{-x^2},
\label{imchi}
\end{equation}
$x = (\omega/k) \sqrt{\beta m/2}$, $\bar\kappa = \kappa a$, $\bar k = ka$. The function $(1/\beta n) \RE \chi(x)$, shown plotted in Fig.~\ref{fig:rechi}, reaches its maximum value $(1/\beta n) \RE \chi(x)|_{\rm max} = 0.2847$ at $x = 1.5$.
%%%%%%%%%%%%%%%%%%%%%%%%%%%%%%%%%%%%%%%%%%%%%%%%%%%%%%%
%%%%%%%%%%%%%%%%%%%%%%%%%%%%%%%%%%%%%%%%%%%%%%%%%%%%%%%
Now, for $\bar k$ fixed, $x$ varies with $\omega$ only. Consequently, for a given assigned pair of values $\left( \Gamma, \bar\kappa \right ) $, the left-hand-side (LHS) term of dispersion relation \eqref{3d-disp}, being independent of $\omega$, would graph as a horizontal line above the $x$ axis in Fig.~\ref{fig:rechi}. Then clearly there can be no solution to \eqref{3d-disp} if the horizontal line is situated above $(1/\beta n) \RE \chi(x)|_{\rm max} = 0.2847$, that is, if 
\begin{equation}
\frac{\bar\kappa ^2 + \bar k^2}{3\Gamma}  \geq \frac{\bar\kappa^2}{3\Gamma} > 0.2847,
\label{3d-cond}
\end{equation}
Thus, for $\bar k > 0$, the condition
\begin{equation}
\Gamma < 1.17 \bar\kappa^2,
\label{3d_Gamma_cond}
\end{equation}
is \emph{sufficient} to guarantee that there can be no solutions to eq.~\eqref{3d-disp}. In the long-wavelength $(k\rightarrow 0)$ limit, which is especially of interest in the present work, condition \eqref{3d_Gamma_cond} become \emph{necessary} as well.

Concentrating on the long-wavelength domain where the dispersion is acoustic, \textit{i.e.} $\omega \propto sk$ ($s$ is the sound speed), we now consider the possible solutions of the dispersion relation \eqref{3d-disp}: those belonging to $x < x_{\rm max}$ are obviously ruled out as possibilities since the derivative $\nicefrac{\partial }{\partial \omega} \RE \chi \left( \mathbf k,\omega \right ) \left. \right |_{\omega = \omega(k)}$ can never be positive definite at $\omega = \omega(k)$ [see eqs.~\eqref{damping},\eqref{smalldamphypothesis},\eqref{imchi}].
Consequently, only the descending portion of the curve in Fig.~\ref{fig:rechi} is relevant, and a more restricted lower bound on the coupling parameter can be estimated by selecting the smallest value of $x$ on the descending portion that will ensure that the Landau damping is sufficiently weak to allow the formation of a viable acoustic mode. Table~\ref{tab:yocp} facilitates this selection.

\begingroup
\begin{table*}
\begin{ruledtabular}
\begin{tabular}{|c|c|c|c|c|}
$x$ &$(\nicefrac{1}{\beta n} )\RE \chi \left(\mathbf k,\omega(k) \right )$ & $(\nicefrac{1}{\beta n} )\IM \chi \left(\mathbf k,\omega(k) \right )$ & $\nicefrac{\partial }{\partial \omega} \RE \chi \left(\mathbf k,\omega \right ) \left. \right |_{\omega = \omega(k)}$ & $\gamma/(x \sqrt2)$ \\
\hline
1.60 & 0.2798 & -0.2192 & -0.1311 & -1.4785 \\
1.70 & 0.2667 & -0.1675 & -0.1832 & -0.7604 \\
1.80 & 0.2484 & -0.1249 & -0.2108 & -0.4657 \\
1.90 & 0.2273 & -0.0911 & -0.2194 & -0.3091 \\
2.00 & 0.2054 & -0.0649 & -0.2144 & -0.2141 \\
2.10 & 0.1839 & -0.0452 & -0.2007 & -0.1518 \\
2.20 & 0.1638 & -0.0308 & -0.1820 & -0.1089 \\
2.30 & 0.1456 & -0.0206 & -0.1614 & -0.0783 \\
2.40 & 0.1295 & -0.0134 & -0.1408 & -0.0561 \\
2.50 & 0.1154 & -0.0086 & -0.1216 & -0.0398 \\
2.60 & 0.1033 & -0.0053 & -0.1044 & -0.0278 \\
2.70 & 0.0928 & -0.0033 & -0.0894 & -0.0191 \\
2.80 & 0.0839 & -0.0020 & -0.0766 & -0.0129 \\
2.90 & 0.0762 & -0.0011 & -0.0659 & -0.0085 \\
3.00 & 0.0696 & -0.0007 & -0.0570 & -0.0054
\end{tabular}
\end{ruledtabular}
\caption{Tabulated values of the real and imaginary part of $\chi$ and Landau damping.}
\label{tab:yocp}
\end{table*}
\endgroup

We observe that the Landau damping diminishes, as it should, with increasing $x$. To be on the safe side, we select $x = 2.2$ as the value beyond which the acoustic mode can be considered to be viable. On this basis, we can state that the realization of a long-lived acoustic mode requires that 
\begin{equation}
\frac{\bar \kappa^2}{3\Gamma} < 0.164,
\end{equation}
resulting in a more restrictive lower bound estimate:
\begin{equation}
 \Gamma > 2.03 \bar\kappa^2 .
\label{3D_res_cond}
\end{equation}
The sound velocity $s$, valid for $x \geq 2.4$, is calculated by solving the dispersion relation (for $x$) resulting from the combination of eq.~\eqref{3d-disp} and the large $x$ expansion of \eqref{rechi}:
\begin{equation}
\frac{1}{\beta n} \RE \chi(x) \approx \frac{w}2 + \frac 34 w^2 + \frac{15}{8} w^3 + \frac{105}{16} w^4 + O(w^5),
\label{w_exp}
\end{equation}
$w = 1/x^2$. The inversion of \eqref{w_exp} combined with \eqref{3d-disp} leads to 
\begin{equation}
\beta m s^2 \approx \frac{1}{y} \left( 1 + 3y + 6y^2 + 24y^3 + 180y^4 + O(y^5)  \right ) ,
\label{3d_sound_series}
\end{equation}
\begin{equation}
y = \frac{\bar k^2 + \bar\kappa^2}{3\Gamma}.
\end{equation}
The series \eqref{3d_sound_series} represents an asymptotic expansion, such that using only the first few terms (the number of terms depends on the value of $y$) provides the best approximation. We have found that for $y = 0.1$ the first three terms in the series suffice. Then 
\begin{equation}
\beta ms^2 \approx \frac{3\Gamma}{\bar k^2 + \bar\kappa^2} \left [ 1 + 3 \frac{\bar k^2 + \bar\kappa^2}{3\Gamma} + 6 \left ( \frac{\bar k^2 + \bar\kappa^2}{3\Gamma} \right )^2 \right ].
\end{equation}
seems to be the best approximation for the Yukawa-Vlasov acoustic dispersion.

At long wavelength $k\rightarrow0$ with $k = 0$, eq.~\eqref{3d_sound_series} simplifies to a purely acoustic dispersion
\begin{equation}
\beta ms^2 = \frac{3\Gamma}{\bar\kappa^2} \left [ 1 + 3 \frac{\bar\kappa^2}{3\Gamma} + 6 \left ( \frac{\bar\kappa^2}{3\Gamma}  \right )^2 \right ].
\label{3d_sound_fin}
\end{equation}
Note that if one sets $\kappa = 0$ at the outset (eq~\eqref{3d-disp}) and then goes to the long wavelength limit thereafter, eq.~\eqref{3d_sound_series} morphs into the familiar Bohm-Gross dispersion relation characteristic of the Coulomb gas \cite{Bohm1949}:
\begin{equation}
\omega^2(k) = \omega_p^2 + 3\frac{k^2}{\beta m},
\end{equation}
where $\omega_p^2 = 4\pi Q^2n/m$.

\begin{figure}[ht]
\centering
\includegraphics[width = .9\textwidth]{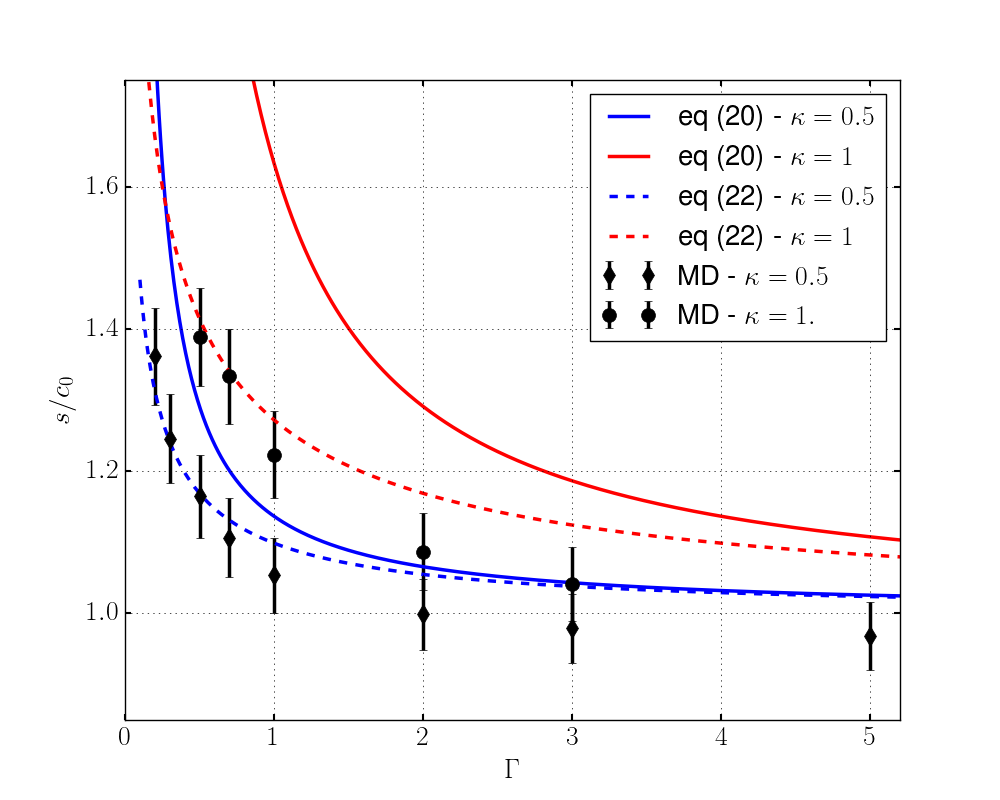}
\caption{Plot of the 3D sound speed obtained from MD simulations and eq.~\eqref{3d_sound_fin} rescaled by $c_0$. The vertical bars indicate a 5\% error.}
\label{fig:MD}
\end{figure}

Our analytic result may now be compared to those of a 3D MD simulations. In Fig.~\ref{fig:MD} we plot the sound speed in units of its $T = 0$ limit, $c_0 = \omega_pa/{\bar \kappa}$, for a range of weak coupling parameters. We notice that in the domain of validity eq.~\eqref{3d_sound_fin} is in reasonable, but not good agreement with the simulations. However, we have also found that the heuristic formula
\begin{equation}
\beta m s^2 = \frac 12 \frac{3\Gamma}{\bar \kappa^2} \left [ 1 + \sqrt{1 + 12 \frac{\bar \kappa^2}{3\Gamma}} \right ] 
\label{luciano_sound}
\end{equation}
- while it represents a poorer approximation to the Yukawa-Vlasov dispersion -  provides a more satisfactory agreement with the MD data. 

The agreement deteriorates for $\Gamma$ values lower than the bound. In fact, simulations show that further reducing $\Gamma$ leads to the disappearance of a well-defined mode \cite{Silvestri2018}. For $\Gamma > 1$ the agreement gets worse again: this is not unexpected, since one leaves the weak coupling zone. The more precise relationship for the weak coupling/strong coupling transition value should be based on \cite{Ott2014}
\begin{equation}
\Gamma_\kappa = \Gamma f(\bar \kappa) = 1,
\end{equation}
\begin{equation}
f(\bar \kappa) = 1 - 0.309 \bar\kappa^2 + 0.0800\bar \kappa^3.
\end{equation}
Thus, the domain of validity for the Vlasov approach to be acceptable is bounded by the condition
\begin{equation}
f^{-1}(\bar\kappa) > \Gamma > 1.17\bar\kappa^2.
\end{equation}
These two conditions are compatible only for $\bar\kappa < 0.82$, further signaling the limitations of the applicability of the Vlasov approach to weakly coupled Yukawa plasmas. 

Finally one may wonder about filling the lacuna created by the absence of the availability of the Vlasov equation below the lower bound. It amounts to finding the appropriate kinetic equation for a weakly coupled gas with a relatively short-range potential. This is still an unsolved problem, although one may suspect that a Boltzmann-like equation may do the job.

%%%%%%%%%%%%%%%%%%%%%%%%%%%%%%%%%%%%%%%%%%%%%%%%%%%%%%%
%%%%%%%%%%%%%%%%%%%%%%%%%%%%%%%%%%%%%%%%%%%%%%%%%%%%%%%
\begin{acknowledgments}
The authors would like to thank Zolt\'an Donk\'o and Peter Hartmann for permission to use their MD simulations data and Marlene Rosenberg for discussions.  This work has been partially supported by NSF Grants No. PHY-­‐0812956, No. PHY-­‐0813153, and No. PHY-­‐1105005.
\end{acknowledgments}
%TC:ignore
%\bibliography{References}

%TC:endignore

\end{document}